# Tuneable terahertz oscillation arising from Bloch-point dynamics in chiral magnets

Yu Li[1*], Leonardo Pierobon[2], Michalis Charilaou[3], Hans-Benjamin Braun[2,4],
Niels R. Walet[5], Jörg F. Löffler[2], James J. Miles[1], Christoforos Moutafis[1,2*]

[1]*Nano Engineering and Spintronic Technologies (NEST) group, Department of Computer Science,
University of Manchester, Manchester M13 9PL, United Kingdom.*
[2]*Laboratory of Metal Physics and Technology, Department of Materials, ETH Zurich,
8093 Zurich, Switzerland.*
[3]*Department of Physics, University of Louisiana at Lafayette, Lafayette LA 70504, USA.*
[4]*Dublin Institute of Advanced Studies, Dublin D04 C932, Ireland.*
[5]*Theoretical Physics, Department of Physics and Astronomy, University of Manchester,
Manchester M13 9PL, United Kingdom.*

Skyrmionic textures are being extensively investigated due to the occurrence of novel topological magnetic phenomena and their promising applications in a new generation of spintronic devices that take advantage of the robust topological stability of their spin structures. The development of practical devices relies on a detailed understanding of how skyrmionic structures can be formed, transferred, detected and annihilated. In this work, our considerations go beyond static skyrmions and theoretically show that the formation/annihilation of both skyrmions and antiskyrmions is enabled by the transient creation and propagation of topological singularities (magnetic monopole-like Bloch points). Critically, during the winding/unwinding of skyrmionic textures, our results predict that the Bloch-point propagation will give rise to an emergent electric field in a terahertz frequency range and with substantial amplitude. We also demonstrate ways for controlling Bloch-point dynamics, which directly enable the tuneability on both frequency and amplitude of this signal. Our studies provide a concept of directly exploiting topological singularities for terahertz skyrmion-based electronic devices.

## I. INTRODUCTION

Magnetic skyrmions are stable topological solitons with unique textures that cannot be continuously unwound into a collinear ferromagnetic state [1]. Various methods of skyrmion formation have been proposed [2,3]. Recent demonstrations of the existence of skyrmions at room temperature [4,5] and current-driven motion in racetrack-based nanostructures [6,7] have made skyrmion-based devices promising candidates for future non-volatile data storage and spintronics-based nanocomputing. Importantly, the study of skyrmionic textures and their formation/annihilation provides a deeper understanding regarding the role of topology in nanomagnetic materials.

In magnetic materials with strong spin-orbit coupling, the competition between antisymmetric Dzyaloshinskii-Moriya interaction (DMI) [8,9] and Heisenberg exchange in conjunction with long-range dipolar interaction generate topologically non-trivial skyrmions. Different classes of DMI arise from different crystal symmetries and therefore a variety of skyrmionic textures can be stabilised, including Bloch [10–12] and Néel skyrmions [5,6]. Moreover, the DMI in materials with $D_{2d}$ crystal symmetry supports antiskyrmions whose existence has recently been theoretically predicted [13,14] and experimentally confirmed [15], either as isolated solitons or in regular lattices. A single skyrmion/antiskyrmion has a quantised topological charge (winding number) [1], which is equal to

$$Q = \frac{1}{4\pi}\int \mathbf{m}\cdot\left(\frac{\partial \mathbf{m}}{\partial x}\times\frac{\partial \mathbf{m}}{\partial y}\right)\mathrm{d}x\mathrm{d}y = \pm 1, \quad (1)$$

with **m** the unit vector of magnetisation. In particular, the sign depends on the polarity, and skyrmions and antiskyrmions of the same polarity can be distinguished classified by the chirality of in-plane magnetisation [13]. In a recent work [16] it has been demonstrated that the magnetic transition of an ideal hexagonal skyrmion lattice involves a transient antiskyrmion lattice state. During that process, the annihilation of skyrmions mediated by a transient state involving antiskyrmions, into a topologically trivial ferromagnetic state has been observed. These magnetic transitions involve complex dynamic variations of the local magnetisation with associated topological winding/unwinding effects.

These dynamic local magnetisation variations induce changes in the electronic state, because electrons adiabatically moving through complex magnetic textures have their spins continuously adjusted to the local magnetisation. Hence, the spatial and temporal variation of the magnetisation will induce emergent electrodynamics of conduction electrons [17,18], which can be represented by an emergent magnetic field **B**$^e$ and an emergent electric field **E**$^e$. In analogy to Faraday's law of induction, for example, a moving skyrmion with a finite velocity will induce an emergent electric field, which directly contributes to Hall measurements as has been experimentally demonstrated by Schulz et al. [19]. During the formation and annihilation of skyrmions, the magnetisation dynamics should also generate dynamic emergent fields; an idea which forms the core of this paper. These fields provide a signal that could allow for the direct electronic detection of skyrmionic switching processes, as well as technological applications of spintronic devices beyond data storage. The overall formation and annihilation processes involve phenomena that manifest topological singularities along with the associated emergent-field considerations, which can have implications from fundamental physics to novel applications in the terahertz range, and thus more detailed studies are essential.

In this work, we start by analysing annihilation phenomena of single skyrmions and antiskyrmions induced by a sweeping external magnetic field, and we show that within numerical models the transition processes and the topology changes are similar for both textures. During these annihilation processes, skyrmion or antiskyrmion tubes break via the creation and subsequent propagation of topological singularities (monopole-like Bloch points). This propagation exhibits a velocity modulation in the terahertz frequency regime, which in turn gives rise to strong high-frequency emergent electric fields. In addition, we propose a design of nanofabricated defects to enable deterministic controllability of the creation and propagation of Bloch points, and thus provide a route to the engineering of robust devices for ultrafast spintronic applications at terahertz frequencies.

## II. RESULTS

### 1. Field-induced annihilation of single skyrmions and antiskyrmions

In order to investigate the intrinsic mechanisms of the transition process from skyrmionic to ferromagnetic states, we focus on the magnetic transitions of single skyrmions and antiskyrmions, thus not considering the interactions with other skyrmionic textures. Our numerical simulations are based on ideal thin films of FeGe with non-centrosymmetric crystal structure [10–12] (see Materials and Methods for the material parameters and simulation methods). Bloch-type skyrmionic configurations can be stabilised numerically in both micromagnetic and atomistic simulations of FeGe by the bulk/intrinsic isotropic DMI. We also find metastable single antiskyrmion tubes in micromagnetic simulations of the same material. The magnetisation in antiskyrmions has different windings along different radial directions away from the centre [13], but bulk DMI only favours either clockwise or anti-clockwise orientation of in-plane spins; in our case the stabilisation of antiskyrmion tubes is expected to arise from contributions of magnetostatic interactions.

In both single skyrmions and antiskyrmions, when a magnetic field of opposite direction to the core polarity is increased in magnitude, the antiskyrmions annihilate at a smaller field than the skyrmions. This is because their more complex spin texture provides additional reversal routes that are not available to skyrmions, but even so, the two processes have similar features (Fig. 1, A and B, and movie S1). If observed in a cross-sectional view (Fig. 1C, taking a single skyrmion tube as an example), the diameter of the skyrmion tube initially decreases, and then unwinds asymmetrically with the core detaching from the bottom surface, followed by the creation of a Bloch point paired with a chiral bobber [20] (Fig. 1D); as the Bloch point propagates through the whole layer towards the top surface, the tube gradually transforms into a trivial uniform ferromagnetic state (Fig. 1E); finally, the entire skyrmion/antiskyrmion tube annihilates when the Bloch point disappears at the top surface (Fig. 1F). It is worth noting that because single skyrmions and antiskyrmions of the same polarity have different winding and opposite topological charge, their annihilation processes are mediated by two different types of Bloch-point singularities [21] (bottom right insets of Fig. 1, A and B). Moreover, the existence of a metastable chiral bobber has been observed in simulations of the annihilation of single skyrmions, either as isolated skyrmions [20] or in skyrmion lattices [12]. Similar magnetic transition phenomena have also been found during the inversion and annihilation in ideal hexagonal FeGe skyrmion lattices (see note S1). However, this is the first observation of antiskyrmionic chiral bobbers.

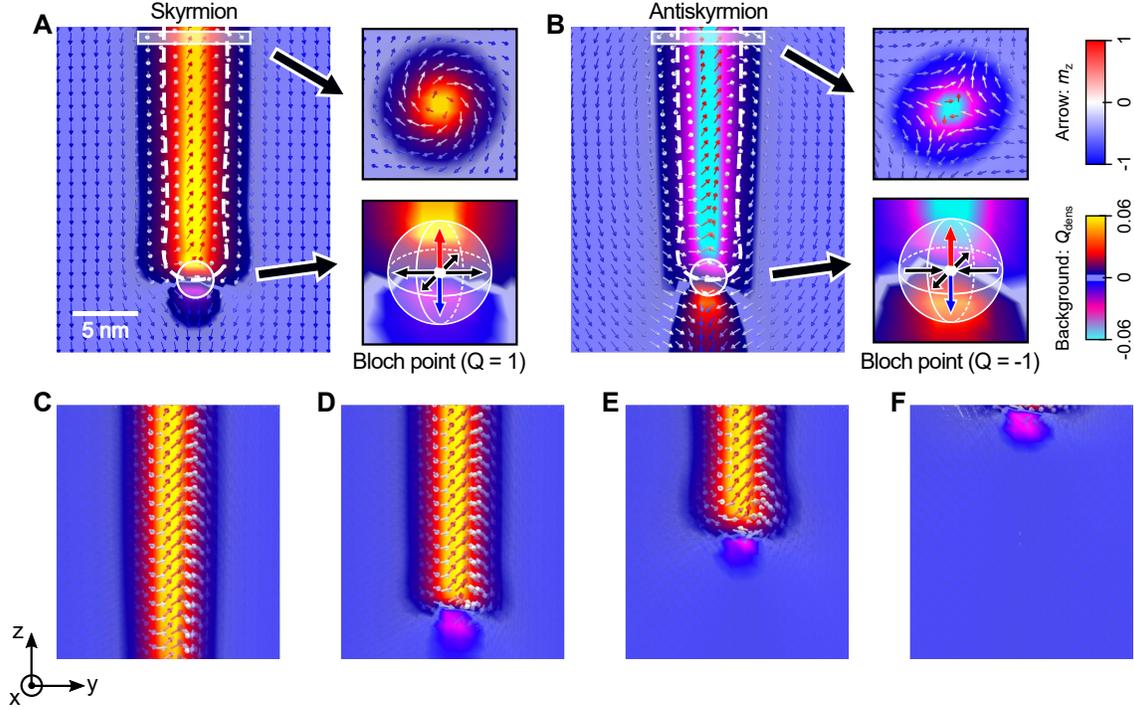

**Fig. 1. Bloch-point dynamics during the field-induced annihilation of single skyrmions/antiskyrmions in FeGe materials.** The magnetisation is shown with arrows and the topological charge density is represented by the background colour. A chiral bobber (dashed-line region) followed by a Bloch point (circled region) is created from a single (**A**) skyrmion and (**B**) antiskyrmion. The top view of magnetisation is shown for a rectangular region in the top-right corner, and a schematic view of the Bloch point is illustrated in the bottom-right corner. (**C** to **F**) Snapshots of the dynamic annihilation process of a single skyrmion computed by atomistic simulations.

## 2. Asymmetry in the creation of Bloch points

As our micromagnetic and atomistic calculations show, a single skyrmion/antiskyrmion tube starts to unwind from the bottom of the film during the annihilation process. Dominated by the isotropic DMI, the magnetisation profile of a single skyrmion located at the centre of the film adopts a Bloch-type configuration. However, the spins near free surfaces experience interactions of neighbours from only one side and thus the effect of short-range interactions is weakened. This induces an inhomogeneous magnetisation profile near the film surface [22]. If the DMI constant is positive the magnetisation will turn outwards near the top surface and inwards near the bottom surface, resulting in a pure Bloch-type skyrmionic structure in the middle of the film, and a mix of a Bloch- and Néel-type skyrmionic structures near the two surfaces (fig. S3, D to F).

As the applied magnetic field is increased in strength, the magnetisation tends to align with the field direction in order to minimise the Zeeman energy, while the precession term makes the in-plane magnetisation precess in the right-handed manner around the field direction, which increases the DMI energy at the bottom surface but decreases it at the top surface (see analytical explanations in the note S2). At a critical field with a magnitude $B_c$, the annihilation process is initiated, and the energy profile at the bottom contributes to a detachment of the skyrmion tube from the surface and, creates a Bloch point. Similarly, if the DMI constant is negative then a Bloch point is created at the top surface and propagates downwards (movie S2).

In order to test the stability of the observed behaviour against finite-temperature effects, we introduced a randomly fluctuating thermal field corresponding to $T = 200$ K (see Materials and Methods), which is below the ferromagnetic Curie temperature of 280 K for FeGe [23]. We find that the annihilation follows a similar process (fig. S5, and Movie S3) as that at zero temperature, and as expected the critical field of the annihilation process is reduced to $B_c = 1.54$ T, because thermal fluctuations contribute to overcoming the energy barrier of the annihilation process.

## 3. Detectable terahertz emergent electric field

As the annihilation process of the skyrmion tube is mediated by Bloch-point dynamics, it involves a rapid change of topological charge: the non-trivial skyrmionic configuration near the surface folds into a topologically equivalent Bloch-point singularities, and the topological charge density rapidly decreases when the Bloch point propagates through the film and leaves a trivial ferromagnetic state in its wake (see Materials and Methods for the calculation of topological charge density). Importantly, the Bloch point is not localised in a single simulation cell because at the centre of the Bloch-point texture the magnetisation vanishes, and the assumption of a continuous magnetisation field breaks down. Hence, for discrete atomistic simulations the centre of a Bloch point has to lie between two atoms, since the magnetic moments of the atoms are assumed to be constant in magnitude. This is also valid in micromagnetic simulations where the Bloch-point centre is between two computational cells.

For single crystals with regular atomic arrangement, such as FeGe [10–12] that we simulated in this paper, the lattice structure induces potential wells between the neighbouring atoms. When a Bloch point propagates through the film, the energy barriers between the potential wells have associated forces that cause the Bloch-point velocity to be modulated. We have performed atomistic simulations to model the dynamics of classical spins and their related magnetic moment in the FeGe system (see Materials and Methods for details of the atomistic simulations). The results show that when a propagating Bloch point experiences three-dimensional periodic potentials, its position/displacement also exhibits corresponding oscillations in the $x$, $y$, and $z$ directions, rather than simply moving upwards with uniform velocity (Fig. 2A, see Materials and Methods for the calculation of Bloch-point position and velocity).

A real-space Berry phase can be regarded as the solid angle subtended by the spin of a conduction electron that is continuously adjusting to the direction of the local magnetisation $\mathbf{m}$ [18]. In a temporally inhomogeneous magnetic texture, the Berry phases result in emergent electromagnetic fields acting on the conduction electrons. The emergent electric field in direction $i = x, y, z$ is given by $E_i^e = \hbar \mathbf{m} \cdot (\partial_i \mathbf{m} \times \partial_t \mathbf{m})$. For the specific case of a Bloch point, which is a non-dimensional topological point defect [1,24], we obtain a quantised topological charge upon integration over the unit sphere $S_2$. When a conduction electron traverses a magnetic texture with a moving Bloch point, it experiences a net force $\mathbf{F} = q\mathbf{E}^e$ with magnitude $F \approx \hbar v_{BP}/2\lambda^2$, where $v_{BP}$ is the velocity of a Bloch point and $\lambda$ is its approximated diameter (we assume $\lambda = 1$ nm in our simulations) [24]. The effect of the emergent electric field on the conduction electrons is thus similar to that of a real electric field. Its strength is linearly dependent on the propagation velocity, and can be easily calculated using the temporal evolution of the Bloch-point position. We find that the velocity $v_{BP}$ is of the order of several kilometres per second and its associated electric field has a magnitude of the order of megavolts per metre (Fig. 2B), similar to the values found in permalloy nanowires as reported by Charilaou et al. [24]. The potential wells originating from the periodic lattice structure discussed above induce rapid oscillations of the Bloch-point velocity and thus the associated emergent field. The fundamental frequency of the emergent electric field extracted with the fast Fourier transform (FFT) method, lies in the terahertz range (Fig. 2C).

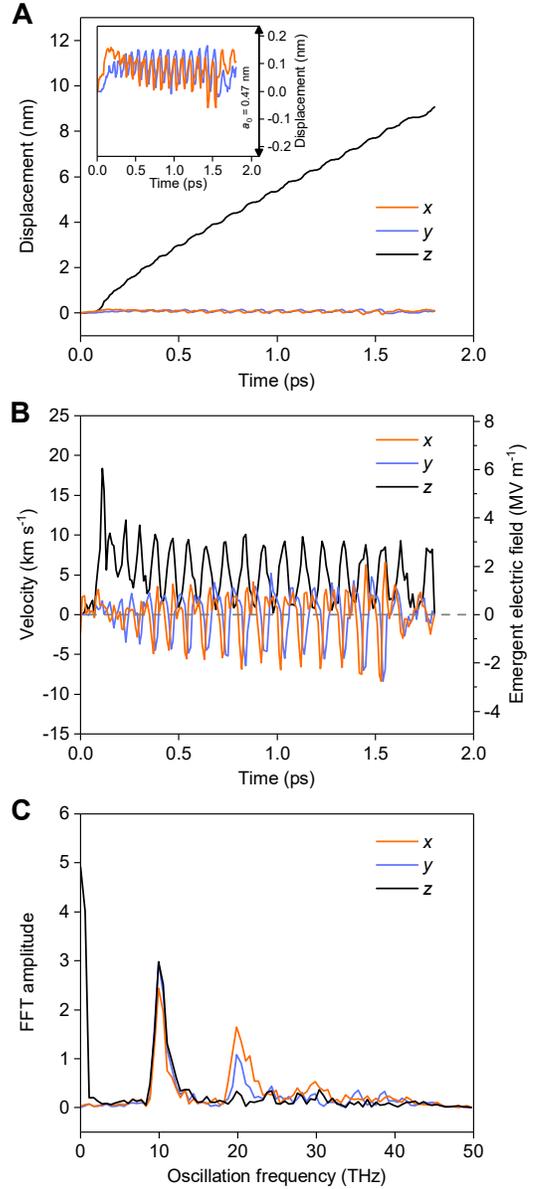

**Fig. 2. Oscillatory Bloch-point motion during a single-skyrmion annihilation.** (**A**) Temporal evolution of the Bloch-point displacement along $x$ (orange line), $y$ (blue line) and $z$ (black line) direction. The displacement along $x$ and $y$ direction are also shown in the enlarged inset. (**B**) Velocity of the Bloch-point propagation with respect to (A), and associated emergent electric field. (**C**) FFT-amplitude spectrum of the velocity oscillation in (B). Note that the peak frequencies are in the terahertz range.

During the final collapse of the structure the Bloch point moves at a velocity of several kilometres per second in a direction perpendicular to the plane of the magnetic layer, but the periodic lattice potential (inset of Fig. 2A) causes oscillations and the consequent emergent field of strength ~2 MV/m is similar in all three dimensions (Fig. 2B). This is different from previous related experimental work [19], which focussed on a study of the electric field arising solely from the in-plane motion of entire skyrmions where the electric current is required to obtain a temporally varied magnetic texture. Here, in contrast, when the skyrmion annihilates via the Bloch-point-mediated mechanism, the radiating emergent electric field is solely excited by a uniform magnetic field. This suggests that the formation and annihilation processes of single skyrmions may be electronically detectable, and it also gives rise to a new mechanism for the generation of terahertz-frequency emergent electromagnetic fields.

In a micromagnetic model, one averages over a large number of atomic spins and thus the Bloch-point behaviour and the resulting emergent field cannot be accurately predicted. However, the mesh-friction effect in micromagnetics [25], where potential wells are induced by finite-size meshes to discretise the magnetisation in practical calculations may cause similar phenomena (see micromagnetic results in the note S3), i.e., the Bloch point exhibits oscillatory motion that can generate an associated electric field (fig. S4, A and B). As the mesh-friction effect that originates from the discrete lattice structure is relevant to the lattice constant, different cell size in micromagnetics compared to the lattice constant may cause the pinning potential to be significantly overestimated/underestimated. Therefore, we studied the effect of different cell sizes in micromagnetic simulations (fig. S4, D and E). In these simulations the values of the fields are of the same order of magnitude as those predicted by the atomistic model, demonstrating that both models are able to produce comparable results. In particular, when the cell size is comparable to the lattice constant $a_0$, the critical fields of the annihilation processes are very similar (1.86 T in the atomistic simulation and 1.82 T in the micromagnetic one), which suggests that the energy barriers in the two models produce comparable effects. Although it is generally difficult to assess the micromagnetic accuracy of the Bloch-point velocity modulation, such models can show behaviour trends when the material parameters are varied. Since micromagnetic simulations are computationally much more efficient than atomistic simulations, especially for the calculation of demagnetisation fields in large systems, we mainly present micromagnetic models for the systematic investigations to reveal more Bloch-point dynamics in the subsequent sections of this paper, assuming qualitatively similar results compared to those obtained from atomistic simulations.

## 4. DMI and thickness-dependent behaviour of Bloch points

As described above, we attributed the asymmetries in the creation of the Bloch point and the annihilation of the skyrmion tube to the distribution of the DMI energy profile through the film driven by the sweeping magnetic field. The surface at which a Bloch point is created is determined by the sign of the DMI in the material, i.e., if the DMI constant is positive, a Bloch point is created at the bottom surface and propagates upwards (Fig. 3A). Conversely, if the DMI constant is negative, a Bloch point is created at the top surface and propagates downwards (Fig. 3B, and movie S2).

The magnetostatic energy plays a crucial role in the stabilisation of skyrmionic textures in three-dimensional systems. In order to investigate the influence of the magnetostatic energy on the position where the Bloch point is created, we varied the film thickness and divided the system into 1-nm-thick layers along the $z$-axis. The corresponding magnetostatic energy of each layer is calculated at the applied field $B_{ext}$ with a magnitude 1.63 T, which is less than $B_c$ for any thickness (fig. S6). The demagnetisation field provides greater stability for a skyrmion structure near the film surface than in the centre of the film, whereas isotropic DMI provides greater stability for a skyrmion in the centre of the film than at its surface [26,27]. Thus, the combined effect of the two interactions results in an inhomogeneous energy distribution throughout the film, where the location of the maximum energy depends on the film thickness.

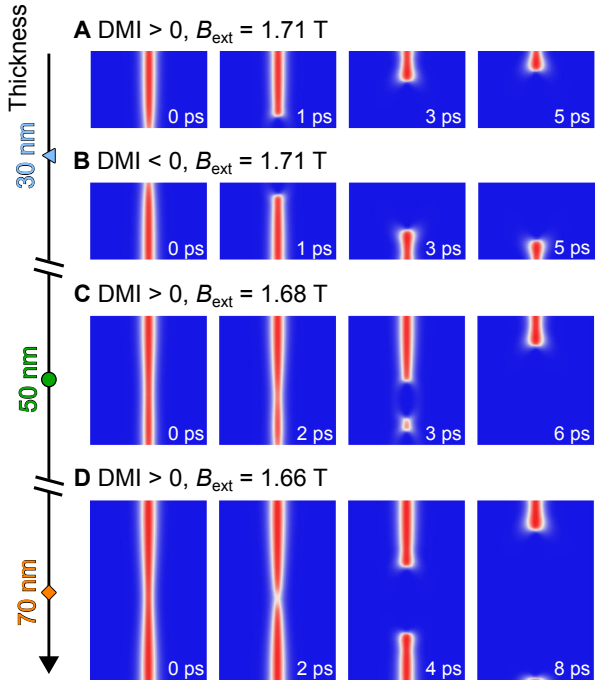

**Fig. 3. Thickness dependence of Bloch-point dynamic processes during skyrmion annihilation.** (**A** and **B**) For a film thickness of 30 nm, the Bloch points emerge from (A) the bottom surface for positive DMI, and (B) the top surface for negative DMI. (**C**) For a thickness of 50 nm a pair of Bloch points is created inside the film together with an individual Bloch point at the bottom. (**D**) For a thickness of 70 nm, a pair of Bloch points is created inside the film.

For a thin film (thickness below 45 nm in our case, around $0.6L_D$ with FeGe helical period $L_D = 70$ nm [28]), the magnetostatic energy inside the film is lower than at the surface, and the Bloch point is created at the bottom surface for positive DMI (Fig. 3A and movie S4) or the top surface for negative DMI (Fig. 3B), as we showed above. For a thick film (thickness above 55 nm in our case, around $0.8L_D$), the energy is significantly higher in the middle than at the surface and a pair of Bloch points emerges close to the middle (Fig. 3D). Then in the intermediate regime (thickness between 45 and 55 nm in our case, around 0.6-$0.8L_D$), the magnetostatic energy in the middle layer is comparable to the energy at the surface, and consequently, the skyrmion tube becomes thinner in the middle and a pair of Bloch points is created close to the middle at the same time as an individual Bloch point emerges at the surface (Fig. 3C). It is worth noting that the Bloch-point pair is not created exactly in the middle of the film, but slightly closer to the bottom surface. This can be attributed to the influence of a positive DMI value, and conversely the Bloch-point pair emerges closer to the top surface for a negative DMI. Similar, but less diverse Bloch-point-mediated switching processes in trivial magnetic bubbles has previously been reported: the bubbles become thinner in the middle of the film where the magnetostatic energy is higher [29], so that the creation of Bloch points always starts in the middle of the film. Here, we demonstrate that the site of the Bloch-point creation is determined by the relative sizes of the thickness-dependent magnetostatic energy and the DMI energy term, setting up a tighter connection of Bloch points and skyrmionic textures in magnetic materials.

## 5. Tuneable Bloch-point dynamics by Gilbert damping and magnetic field

Since Bloch-point creation and propagation are a series of dissipative dynamic magnetisation phenomena, their behaviour depends on the Gilbert damping parameter α, which is known to slow down the depinning of magnetic domain walls [30]. In this paper, we use α = 0.3 for FeGe [31], but we have also analysed these phenomena for a wider range of α to look at the universality of our results (Fig. 4, A and B, and fig. S7, A and C). Particularly noteworthy are the cases of ultra-low damping, α = $6\times10^{-5}$, where we find a large amplitude for the oscillatory phenomena on the skyrmion tube during the Bloch-point propagation (Fig. 4A), and that of critical damping, α = 1, where we observe a direct energy-minimisation process (Fig. 4B). The value α = $6\times10^{-5}$ is comparable to the damping parameter of sub-100-nm-thick Yttrium Iron Garnet (YIG), which has the lowest known damping of any thin film [32]. For a low damping (α < 0.1 in our case), the amplitude of the magnetisation oscillation before the Bloch point approaches the top surface is sufficiently large to detach the skyrmion tube from the top surface. This leads to the creation of a new Bloch point (α = 0.01, from 2nd to 3rd panel in fig. S7A) below the top surface, or even a Bloch-point pair for an ultra-low damping (α = $6\times10^{-5}$, from 2nd to 3rd panel in Fig. 4A).

We find that the Bloch-point propagation velocity increases with increasing α and saturates close to the critical damping (Fig. 4D). This is to be expected since increasing the damping increases the rate at which the magnetisation moves towards the minimum energy in the Landau-Lifshitz equation. This dependence of velocity on the damping parameter is opposite to the one observed in a current-driven skyrmion longitudinal motion along a racetrack [33], where the velocity is inversely proportional to α due to a different mechanism. It is worth noting that in the low damping region (α < 0.1), complex processes such as the creation of additional Bloch points or pairs do not significantly affect the average velocity.

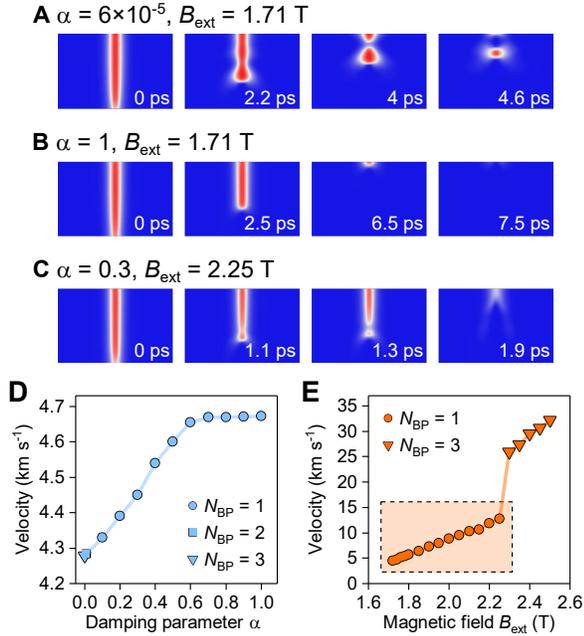

**Fig. 4. Effect of Gilbert damping and external magnetic field on the Bloch-point behaviour.** (**A** to **C**) The number of Bloch points that are created and propagate depends on α and $B_{ext}$. (**D** and **E**) Dependence of the Bloch-point velocity on (D) the damping parameter at $B_{ext}$ = 1.71 T and (E) the external magnetic field at α = 0.3. The number of Bloch points $N_{BP}$ is indicated by the symbols as shown in the legends. The solid lines connecting the data points are guides to the eye.

The strength of the applied magnetic field affects the Bloch-point propagation velocity more straightforwardly and has a greater effect than damping (fig. S7, D and E). Whereas the velocity between the cases of ultra-low damping and critical damping increases by only 10%. Figure 4E shows that for α = 0.3 the propagation velocity linearly increases with the magnetic field strength and can be doubled by applying a field that is around 15% higher than $B_c$. It is worth noting that at around $B_{ext}$ = 2.25 T, there is a sudden velocity jump attributed to another Bloch-point pair breaking off from the chiral bobber as the skyrmion tube unwinds (Fig. 4C). Interestingly, when a high magnetic field, $B_{ext}$ = 1.80 T, is applied in the ultra-low damping case, two Bloch-point pairs are created, i.e., one Bloch-point pair during the propagation due to the application of the field, and another pair just below the top surface similar to the behaviour of the ultra-low damping case (fig. S7C, and see all the above-mentioned dynamic processes in the movie S5).

The Bloch-point velocity can be modulated by applying different strength of magnetic field, and the oscillation frequency of the emergent electric field is determined by the oscillatory velocity of the Bloch point when it propagates through a tilted 'washboard' potential originating from discrete crystal lattices (Fig. 5, A and B). Therefore, as the applied field increases so does the Bloch point velocity and the associated frequency. The relationship is illustrated in Fig. 5C, which shows a linear dependence. Our findings, therefore, show that the Bloch-point propagation during the annihilation of skyrmions will result in a terahertz-level emergent electric field with a duration of several picoseconds, and the frequency of which can be controlled by an external magnetic field, prompting us to envisage possible applications in future spintronics devices.

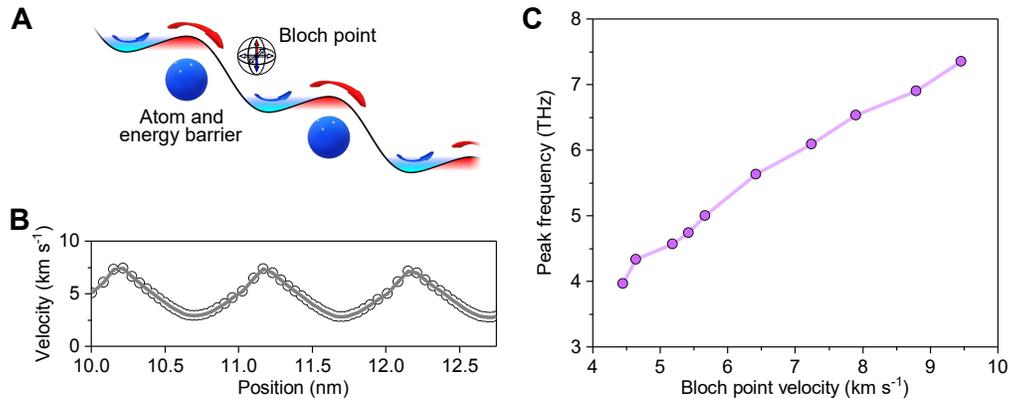

**Fig. 5. Dependence of the peak-oscillation frequency on the Bloch-point propagation velocity.** (**A**) Schematics of the Bloch-point propagation with a tilted 'washboard' potential originating from discrete crystal lattices. This results in (**B**) oscillation of the propagation velocity when a Bloch point is hopping through the film. (**C**) Peak frequency of the velocity as a function of the Bloch-point propagation velocity. The solid lines connecting the data points are guides to the eye.

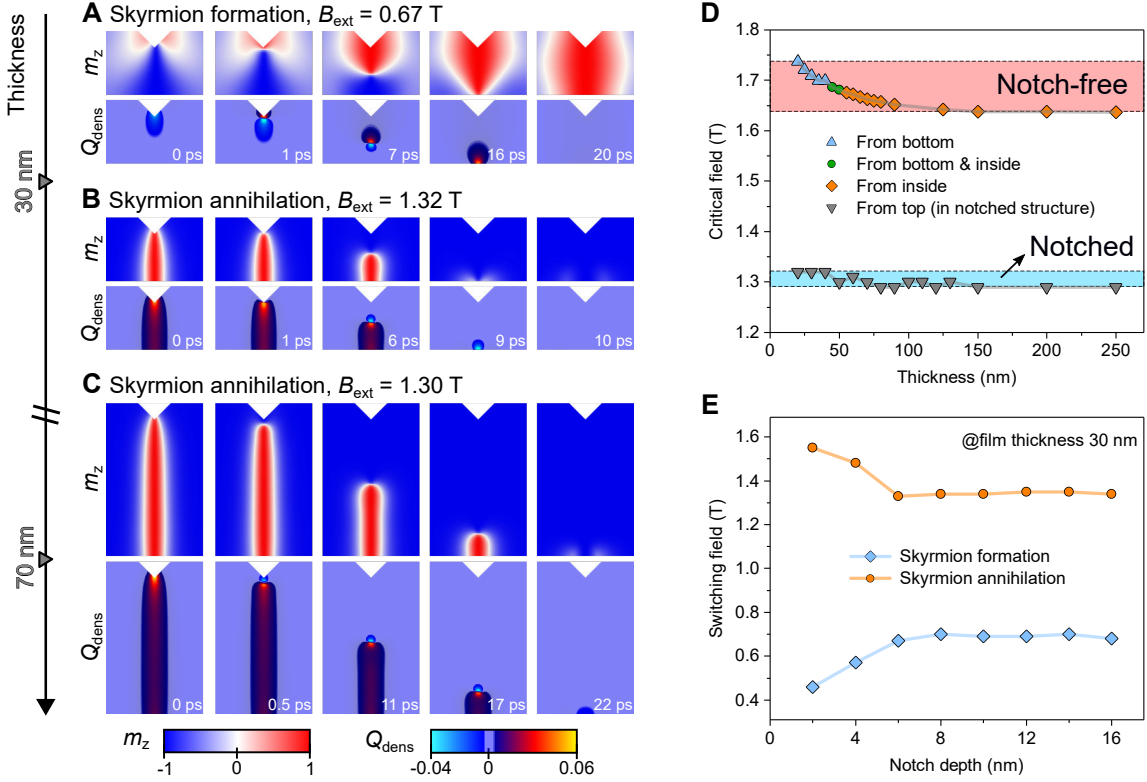

**Fig. 6. Deterministic formation and annihilation of a single skyrmion by an artificial notch.** (**A**) Skyrmion formation occurs at $B_{ext}$ = 0.67 T as the magnetic field is swept from 1.50 T to 0.50 T. (**B**) Skyrmion annihilation occurs at $B_{ext}$ = 1.32 T as the field is swept downwards. (**C**) With the presence of a notch the skyrmion starts to annihilate from the top surface at $B_{ext}$ = 1.30 T for a film thickness of 70 nm. (**D**) Dependence of the critical field of annihilation processes on layer thickness with notch-free (light red band) and notch-based (light blue band) structure. (**E**) The dependence of the switching (formation or annihilation) field on notch depth (for a film thickness of 30 nm). Solid lines serve as a guide to the eye.

## 6. The effect of an engineered notch

Racetrack systems have previously been designed with in-plane non-magnetic notch structures to enable geometrically localised disordered spin textures and control the lateral position of domain walls, which may result in domain wall pinning [34] or skyrmion formation [35] when driven by electric currents. Here, we propose using three-dimensional notch structures to realise the manipulation of Bloch-point dynamics, since the site of the Bloch-point creation can be effectively determined by the notch structure. In this way, we propose a new mechanism where the creation of Bloch-point singularities as well as the writing and deleting of individual skyrmions can be achieved by applying a magnetic field in an appropriate strength.

We have simulated the effect of a conical notch near the top surface. The notch has a diameter of 10 nm and a depth of 5 nm, and the saturation magnetisation $M_s$ was set to zero inside the notch region (Fig. 6). To determine the effect of the notch we studied the behaviour of skyrmions in a uniform magnetic field, starting from the uniform ferromagnetic state, and then swept the external field from $B_{ext}$ = 1.50 T to 0.50 T. The artificial defect changes the geometry of the system, effectively modifying the demagnetisation field surrounding the notch region and consequently changing the distribution of magnetostatic energy density throughout the film. The local magnetostatic fields cause the spins to twist around the non-magnetic notch, causing an accumulation of topological charge density around the notch which lowers the energy barrier for the creation of a Bloch point. In Fig. 6, we plot both the local magnetisation ($m_z$) and the topological charge density ($Q_{dens}$) to illustrate the dynamic winding and unwinding of the spin textures around the notch structure. At $B_{ext}$ = 0.67 T, a Bloch point is created below the notch tip, and as the Bloch point propagates downwards, a chiral bobber is generated and merges with the propagating Bloch point until it reaches the bottom surface, where the Bloch point disappears and a complete skyrmion forms (Fig. 6A).

Page **8** of 12

Importantly, the skyrmion that has been formed can be annihilated if the field is increased again to 1.50 T (Fig. 6B). The skyrmion becomes thinner as the magnitude of the field increases, and at $B_{ext}$ = 1.32 T it detaches from the notch and creates a Bloch point, which propagates downwards until the whole skyrmion tube finally switches back to the trivial ferromagnetic state. The global magnetisation and topological charge are monitored during the whole notch-induced skyrmion formation and annihilation processes (fig. S8, and movie S6). The results show that even in a material with a positive DMI, Bloch points will emerge at the tip of the notch near the top surface, in contrast to the creation at the bottom surface in a notch-free system. Hence, with a notch structure we can tune the balance between DMI and magnetostatics.

Even though the notch depth (5 nm) is much smaller than the thickness of the films that we have studied (20-250 nm), the modifications to the magnetisation profile and associated magnetostatic energy are substantial. The rotational magnetisation around the notch enables annihilation to occur at significantly lower fields than in films without the notch. Taking the 70-nm-thick sample as an example, the skyrmion tube starts to annihilate at $B_c$ = 1.66 T in the notch-free system (Fig. 3D), but at $B_c$ = 1.30 T from the top side in the notch-based system (Fig. 6C). Meanwhile, at this thickness the notch-free film would normally create a Bloch-point pair inside the film resulting in a complex emergent electric field with two superimposed oscillation frequencies, as the two Bloch points will propagate in opposite directions with slightly different velocity in the initial stage (fig. S9A). In contrast, the notch generates only a single Bloch point that propagates through the film, giving rise to an emergent field with a more uniform frequency and amplitude (fig. S9B). We have also varied the film thickness to show that the critical field of annihilation processes in the notch-based system is almost independent of the thickness, i.e., the variation in the field strength (light blue region in Fig. 6D) is much smaller with a notch compared to the notch-free structure (light red region in Fig. 6D).

In order to show that these phenomena can be mainly attributed to the distortion of the spin texture around the notch, we have also investigated how the depth of the notch structure affects the switching fields for the formation and annihilation of skyrmions. Figure 6E shows that at small notch depths, the surface charge and a resulting magnetostatic field have an influence on the magnetostatic energy distribution [36], which weakens the notch-induced localised distortion. However, the local distortion inside the system does not vary significantly if the notch depth is large enough (greater than 6 nm in our case), and the switching fields become almost independent of the notch depth. These findings show that relatively deep notches will provide sufficient stability for the control of Bloch-point creation as well as skyrmion writing/deleting in order to make the fabrication robust against engineering errors in the notch size.

## III. CONCLUSION

In conclusion, we have identified the mechanisms of skyrmion annihilation in chiral magnetic thin films with isotropic Dzyaloshinskii-Moriya interaction. Below a certain film thickness, skyrmions are stable and continuous throughout the film. Upon application of an external field, a skyrmion tube breaks and terminates in the form of Bloch points that propagate below the film surface. In thicker films, the skyrmions break inside the film and pairs of Bloch points are created that move in opposite directions towards the surfaces of the film. In both cases, the Bloch points move at speeds of kilometres per second. Surprisingly, the speed is modulated at terahertz frequencies due to the rapid hopping of the Bloch point along the energy wells of the discretised lattice. The Bloch-point motion also generates emergent electric fields with a substantial and consequently measurable magnitude in the megavolt per metre range.

Our parametric studies also reveal the underlying mechanisms of Bloch-point dynamics, which are attributed to the competition between DMI, magnetostatic energy, and external magnetic field strength. We demonstrate that the frequency of the emergent electric field is proportional to Bloch-point propagation velocity, which in turn depends on the external magnetic field. We propose that nanofabricated defects can deterministically localise topological charge, allowing us to tailor the creation and propagation of Bloch points, thus realising controllable generation of the high-frequency electromagnetic signal. Since Bloch-point singularities are the smallest topological defects in spintronic systems, it is tantalising to envisage their possible use in future electronic devices that exploit their emergent intriguing dynamics.

# IV. MATERIALS AND METHODS

## 1. Simulation method

The dynamic micromagnetic simulations were carried out using Mumax$^3$, a finite-difference GPU-accelerated package [37], whose underlying physics is based on the Landau-Lifshitz equation of motion [38]:

$$\frac{\partial \mathbf{m}}{\partial t} = \frac{\gamma}{1+\alpha^2} \left\{ \mathbf{m} \times \mathbf{H}_{\text{eff}} + \alpha \left[ \mathbf{m} \times \left( \mathbf{m} \times \mathbf{H}_{\text{eff}} \right) \right] \right\}, \quad (2)$$

where $\mathbf{m}$ is the magnetisation unit vector, $\gamma$ is the gyromagnetic ratio of the electron, and $\alpha$ is the dimensionless Gilbert damping parameter. Here, we used $\alpha = 0.3$ [31] if not specified otherwise. Furthermore, $\mathbf{H}_{\text{eff}}$ is the effective magnetic field experienced by the magnetisation $\mathbf{M} = M_s \mathbf{m}$ with constant magnitude $M_s$, and is defined as $\mathbf{H}_{\text{eff}} = -\mu_0^{-1} \delta E / \delta \mathbf{M}$ with the vacuum permeability constant $\mu_0$. The energy density $E_v$ is written as:

$$E_v = A(\nabla \mathbf{m})^2 + D\mathbf{m}(\nabla \times \mathbf{m}) - K_{u1}(\mathbf{n} \cdot \mathbf{m})^2 - \mu_0 M_s \mathbf{H} \cdot \mathbf{m} - \frac{\mu_0 M_s}{2} \mathbf{H}_d(\mathbf{m}) \cdot \mathbf{m}, \quad (3)$$

which is contributed from: (1) a symmetric exchange interaction with exchange stiffness $A$; (2) an isotropic Dzyaloshinskii-Moriya interaction (DMI) for crystallographic classes of $D_n$ symmetry as a sum of Lifshitz invariants [39] with a strength $D$; (3) a magnetocrystalline anisotropy with the first-order uniaxial anisotropy constant $K_{u1}$ and its direction $\mathbf{n}$; (4) a Zeeman coupling of the magnetisation with the external magnetic field $\mathbf{H}$; and (5) a magnetostatic interaction energy of the magnetisation in the field $\mathbf{H}_d$ generated by volume and surface magnetostatic charges.

In our micromagnetic simulations, Eq. 2 is solved using the Dormand-Prince method (RK45 solver) with adaptive time steps. The magnetic parameters were adapted from the cubic FeGe chiral magnets [31], with exchange stiffness $A = 8.78$ pJ m$^{-1}$, DMI strength $D = 1.58$ mJ m$^{-2}$, and saturation magnetisation $M_s = 384$ kA m$^{-1}$. Bulk FeGe has an isotropic lattice structure and consequently no magnetocrystalline anisotropy. Although the existence of anisotropy has been experimentally validated in epitaxial thin films [12,40], our initial simulations indicate that the experimentally observed values do not produce qualitatively different changes on the simulated behaviour of skyrmions. Furthermore, because the anisotropy of thin films is relevant to the substrate and deposition methods, we used the bulk values. The overall system was divided into 1 nm × 1 nm × 1 nm cubic cells. Infinite samples were simulated by setting the dimensions in the *xy*-plane to 200 nm × 200 nm, with periodic boundary conditions in the plane, and in the *z*-direction the default thickness is 30 nm unless specified otherwise.

The atomistic methods are based on the open-source spin simulation framework — Spirit [41]. The ideal crystalline B20 FeGe is a simple cubic lattice structure with a lattice constant $a_0 = 0.47$ nm and four Fe atoms in the $P2_13$ space group, in Wyckoff position (*x*, *x*, *x*), (-*x*+1/2, 1-*x*, *x*+1/2), (-*x*, *x*+1/2, -*x*+1/2), and (*x*+1/2, -*x*+1/2, -*x*), with *x* = 0.3865 [42]. We calculated the exchange coupling between neighbouring Fe atoms necessary to achieve consistency with the micromagnetic model as $J = Aa_0/n = 6.45$ meV/pair, where n is the number of atoms per unit cell [43]. We took $n = 4$ and only considered the first-nearest-neighbour exchange interactions. The DMI strength between first-nearest-neighbour spins is $D_a = 2\pi a_0 J/L_D = 0.27$ meV/pair, where the helical period of FeGe $L_D$ is 70 nm [28]. The local spin moment of Fe is $\mu_s = M_s a_0^3 / n = 1.08 \mu_B / \text{atom}$, similar to the theoretical value of $1.16 \mu_B$/atom [44], with $\mu_B$ the Bohr magneton [43]. The atomistic simulation system was 50 × 50 × 20 unit cells (thickness 9.4 nm) with periodic boundary conditions in the *x* and *y* (in-plane) directions.

## 2. Finite-temperature simulation.

The stability of our results against non-zero temperatures is tested by introducing a randomly fluctuating thermal field to $\mathbf{H}_{\text{eff}}$, and is determined by W. F. Brown [45]:

$$\mathbf{H}_{\text{therm}} = \boldsymbol{\eta} \sqrt{\frac{2\alpha k_B T}{\mu_0 M_s \gamma V \Delta t}}, \quad (4)$$

with a randomly oriented normal vector $\boldsymbol{\eta}$, the Boltzmann constant $k_B$, the temperature $T$ (200 K in our simulations), the volume of single computational cell $V$ and the time step $\Delta t$. The simulations are implemented using the 6$^{\text{th}}$ order Runge-Kutta-Fehlberg (RKF56) solver with adaptive time steps [46].

## 3. Calculation of topological charge density

The topological charge denotes the winding of spin texture in a mapping $S^2 \rightarrow S^2$, and the 2-sphere $S^2$ can be either the compactified plane $R^2$ or a sphere surrounding a point [1]. For a 3D skyrmion tube, it has a quantised charge $Q = 1$ in each slice of the *xy*-plane perpendicular to the film. For an illustration of the topological-charge evolution during the annihilation process of a skyrmion/antiskyrmion tube, we plotted the topological charge density

$Q_{\text{dens}} = \frac{1}{4\pi} \mathbf{m} \cdot (\partial_x \mathbf{m} \times \partial_y \mathbf{m})$. In our discrete micromagnetic and atomistic system, the equation is implemented with a MATLAB script, using a higher-order polynomial interpolation to approximate the partial derivatives $\partial_x \mathbf{m}$ and $\partial_y \mathbf{m}$. The lattice version of topological charge introduced by Berg et al. [47] has also been implemented by calculating the solid angle from each neighbouring triangle magnetisation group. The results obtained from both methods exhibit a high level of consistency, confirming the accuracy of our calculations.

### 4. Calculation of Bloch-point position and velocity

For the calculation of the Bloch-point propagation velocity, the localisation of the Bloch point is of vital importance. In our simulation, we calculate the Bloch-point centre using the method proposed by Thiaville et al. [25], in which the location of the Bloch point is determined by a polynomial interpolation of the magnetisation from nearby mesh points.

### Acknowledgments

This work was supported by the Royal Society International Exchanges programme (Ref: IE161506) and by the Swiss National Science Foundation (Grant No. 200021-172934). Y.L. acknowledges a scholarship supported by the China Scholarship Council.